\documentclass[aps,twocolumn,superscriptaddress,prl,amsmath,amssymb,10pt,aps,longbibliography]{revtex4-1} 
\usepackage[colorlinks=true]{hyperref}
\usepackage{lipsum}
\usepackage{graphicx}
\usepackage{latexsym}
\usepackage{amsmath}
\usepackage{amsthm}
\usepackage{amssymb}
\usepackage{epstopdf} 
\usepackage{enumitem}
\usepackage{setspace}
\usepackage{dcolumn}
\usepackage{bm}
\usepackage{setspace} 
\usepackage{slashed}
\usepackage{color}
\usepackage{youngtab}
\usepackage{tikz}
\usepackage{tikz-3dplot}
\usepackage{braket}
\usepackage{makeidx}
\usepackage{cancel}
\usepackage{subfigure}
\usepackage{pbox}
\usepackage{kotex}
\usepackage[normalem]{ulem}

\usepackage{lipsum}
\graphicspath{{Figures/}}

\usetikzlibrary{shapes,snakes,arrows,chains,matrix,positioning,scopes,calc}
\usetikzlibrary{decorations.markings}

\usepackage{moreverb}
\allowdisplaybreaks 
\begin{document}
	
	
	\title{Dynamic orders of a Quantum Spin Liquid at Non-zero Temperatures}

	\author{Minsu Park}
	\affiliation{Department of Physics, Korea Advanced Institute of Science and Technology (KAIST), Daejeon 34141, Korea}
	
	\author{Masafumi Udagawa}
	\thanks{Masafumi.Udagawa@gakushuin.ac.jp}
	\affiliation{Department of Physics, Gakushuin University, Mejiro, Toshima-ku 171-8588, Japan}

	\author{Eun-Gook Moon}
	\thanks{egmoon@kaist.ac.kr}
	\affiliation{Department of Physics, Korea Advanced Institute of Science and Technology (KAIST), Daejeon 34141, Korea}

	\date{\today}
	\begin{abstract}
	A quantum spin liquid hosts massive quantum entanglement whose identification is one of the most significant problems in physics. Yet, its detection is known to be notoriously difficult because of featureless properties without a symmetry order parameter. 
	Here, we demonstrate dynamic signatures of a quantum spin liquid state by investigating Kitaev's spin model on the hyper-honeycomb lattice, where a quantum spin liquid state is stabilized as a stable thermodynamic phase. 
	The real-time dynamics of spin correlation function is obtained with the large-scale quantum Monte Carlo simulation.
	We find the onset of a characteristic oscillation in dynamic local spin correlation as entering the  quantum spin liquid phase.
	Our results show that a quantum spin liquid may be characterized by a sharp growth of coherent spin dynamics of the system, which we name as a dynamic order.   
	We further propose that a dynamic-order may naturally detect a featureless thermal phase transition, which has been reported in a class of strongly correlated materials. 
	\end{abstract}

	\maketitle

	\textit{	Introduction.\textbf{---}} 
	Massive entanglement of quantum many-body systems may intrinsically appear in quantum spin liquids which has been proposed as one of the main states in future science and technology \cite{Zhou2017.4,Savary2016.11,Knolle2019.3}.  The spin model on the honeycomb lattice proposed by Kitaev  realizes such a quantum spin liquid where any  magnetic orderings are prevented by the intrinsic massive entanglement of quantum spins \cite{Kitaev2005.10}.

	An amount of theoretical and experimental works have been performed in literatures.
	Analytic and numerical studies with density matrix renormalization group, quantum monte carlo (QMC) and exact diagonalization have provided the insights of physical observables including thermal conductivity, specific heat, and spin structure factor \cite{Plumb2014.7, nasu2014vaporization, Koitzsch2016.9,Sandilands2015.4,Kim2015.6,Kim2016.4, smith2016majorana,knolle2014dynamics,knolle2015dynamics,yoshitake2017majorana,yoshitake2016fractional,yoshitake2017temperature,yoshitake2020majorana,gohlke2018quantum,gohlke2018dynamical,gohlke2017dynamics, Winter2017.11,Winter2016.6,Winter2018.2,Yadav2016.11,Takeda2022.11,Viciu2007.1,Perkins2017,Starykh2018,udagawa2018vison,udagawa2019spectroscopy, PerkinsPRB2019,Songvilay2020.12,Lin2021.9,Wulferding2020.3,Tanaka2022.1,wang2021fractionalized}. The experimental progress has been made in identifying and characterizing candidate materials, such as  $\alpha$-RuCl$_3$\cite{singh2010antiferromagnetic,singh2012relevance,plumb2014alpha,sandilands2015scattering,banerjee2016proximate,kitagawa2018spin,roudebush2016iridium,abramchuk2017cu2iro3,
	bahrami2019thermodynamic,banerjee2017neutron,sears2015magnetic,johnson2015monoclinic,chaloupka2013zigzag,rau2014generic,banerjee2018excitations,majumder2015anisotropic,kasahara2018majorana,yamashita2020sample,yokoi2021half,wulferding2020magnon,ponomaryov2017unconventional,wang2017magnetic,yokoi2021half,czajka2021oscillations,zhang2021topological,bruin2022robustness,lefranccois2022evidence,czajka2023planar,koitzsch2017nearest,zhou2019evidence,mashhadi2019spin,weber2016magnetic,gronke2018chemical,zhou2019possible,carrega2020tunneling,feldmeier2020local,konig2020tunneling,udagawa2021scanning,takahashi2023nonlocal,nagai2020two,hwang2022identification,go2019vestiges,imamura2024majorana,noh2024manipulating,choi2020theory,namba2024two,tanaka2022thermodynamic}, and a trait of Majorana fermions at non-zero temperatures has been analyzed by assuming that its signature has survived at non-zero temperatures \cite{PerkinsPRB2019}.

At non-zero temperatures, strictly speaking, quantum spin liquids are only stable in three spatial dimensions while they become unstable in two spatial dimensions due to strong gauge fluctuations and their confinement in two spatial dimensions. In particular, in three-dimensional $Z_2$ spin liquids, a thermal phase transition between a quantum spin liquid and a trivial paramagnetic state exists without any symmetry order parameters.  The transition is caused by the condensation of string-type excitations, accompanying the release of large entropy around the critical point. The character of thermal transition is clearly described by the exact solution for the Toric code model,  
and as observed for the three-dimensional Kitaev model in numerical calculations with QMC \cite{nasu2014vaporization}. Note that a class of strongly correlated systems including Sr$_2$VO$_3$FeAs \cite{KimExp2017} have been reported to show such a symmetric thermal transition with a specific heat jump.

In this work, we report an intriguing characteristic of quantum spin liquids at non-zero temperatures by using both numerical and analytical calculations with a spin model on the hyper-honeycomb lattice. We show that dynamic responses of spin observables such as local spin correlation function or spin structure factor may identify quantum spin liquids at non-zero temperatures and their transition temperature to a trivial paramagnetic state. The identifications are associated with symmetric and dynamical responses of the spin observables in drastic contrast to spontaneous symmetry breaking phenomena. Specifically, we uncover emergent behaviors of spin observables including the appearance and disappearance of a oscillation below and above the transition temperature in the local spin correlation function by performing both QMC calculations with finite size systems and analytical calculations with infinite size systems at zero temperature. Thus, we argue that the dynamic and symmetric responses may identify quantum spin liquids at non-zero temperatures, called dynamic orders of quantum spin liquids.

	\begin{figure}[t]
		\includegraphics[width=3cm, height=3cm]{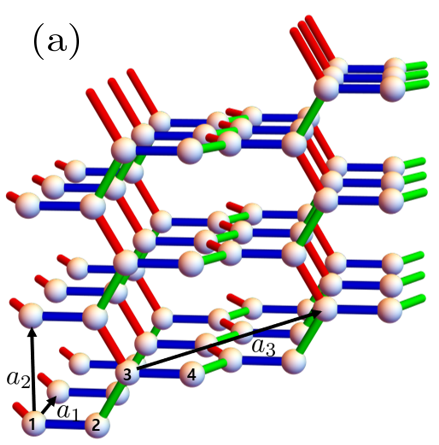}
		\raisebox{0pt}[0pt][0pt]{\hspace{0.5cm}\includegraphics[width=4.5cm, height=3cm]{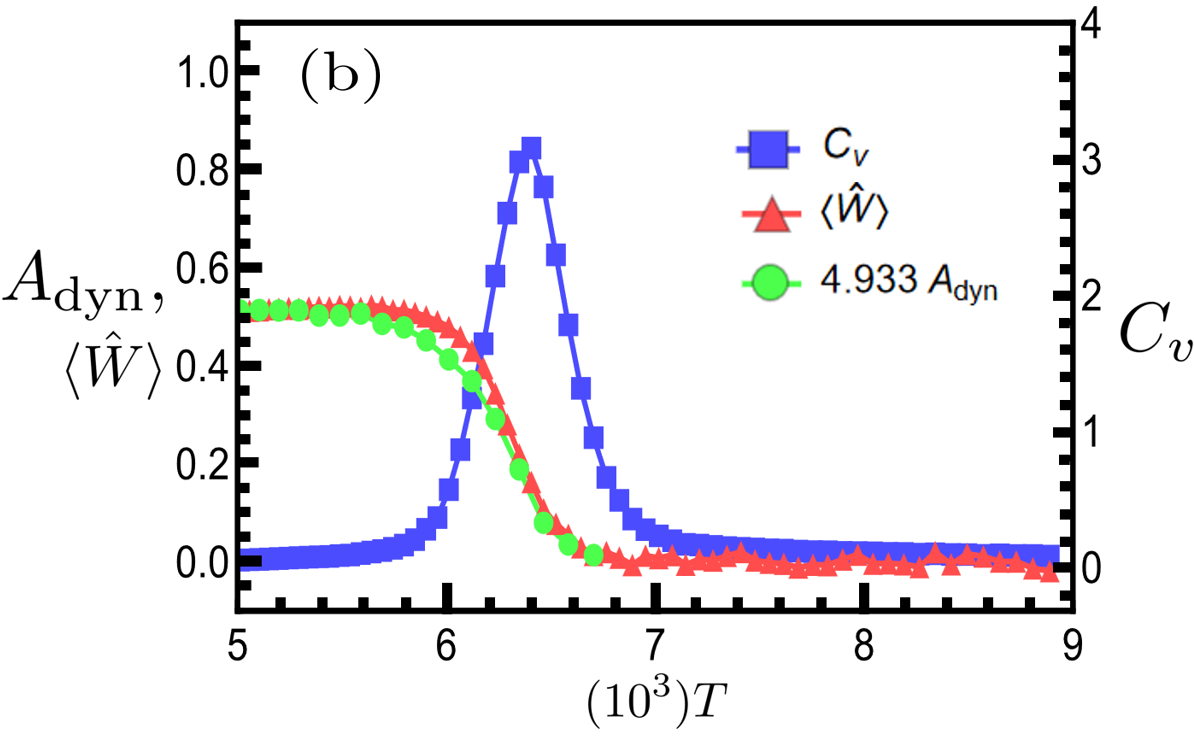}}
		\caption{(a) the graphical representation of the Kitaev model on hyperhoneycomb lattice with $3 \times 3 \times 2$ unit cells. The blue, red, and green bonds are for spin directional exchange interaction terms. (b) Observables for the thermal transition at $T=T_w$. The specific heat (blue) and the expectation value of the non-local Wilson loop operator $\hat{W}$ (red) are shown. The dynamic order ($A_p$) from a {\it local} spin correlation function (green) captures the thermal transition.  } 
	\end{figure}

	\textit{The Model.\textbf{---}}  We consider a spin Hamiltonian on a hyper-honeycomb lattice with $4N^3$ sites and positive energy $J_{x,y,z}$,  
	\begin{align}
		H = - \sum_{\langle ij \rangle_{\gamma}} J_{\gamma}\sigma^\gamma_i\sigma^\gamma_j, \quad \gamma \in {x,y,z} 
	\end{align}
	where $\langle ij \rangle_{\gamma}$ are for the nearest-neighbor bonds and $\sigma^{x,y,z}_i$ are the Pauli operators at a site $i$. This model is an extension of the honeycomb model by Kitaev to a hyper-honeycomb lattice where the three types of the links ($x,y,z$) are well defined as illustrated in Fig.1 (green, blue, red). 
	For the semi-open boundary condition (periodic along $a_3$ and open along $a_1$ and $a_2$), this Hamiltonian can be further simplified by introducing the Majorana representation \cite{nasu2014vaporization} with the Jordan-Wigner transformation,  
	\begin{align}
		H &= iJ_x\sum_{\text{x bonds}}c_wc_b - iJ_y\sum_{\text{y bonds}}c_bc_w - i J_z\sum_{\text{z bonds}}\mu_rc_bc_w \nonumber
	\end{align}
	where the $Z_2$ operator $\mu_r$ (whose eigenvalues are $\pm1$) is defined on each $z$ bond and commutes with the Hamiltonian. This Hamiltonian becomes
	\begin{align}
		H &= \frac{i}{4}\sum_{k,k'}c_kA_{kk'}c_{k'}
	\end{align} 
	after fixing a value of $\mu_r$. The matrix $A$ has a fixed value for $x$ and $y$ bonds, and the sign for $z$ bonds is determined by the fixed value of $\mu_r$.  For periodic boundary conditions, the Hamiltonian also becomes $\frac{i}{4}\sum_{k,k'}c_kA_{kk'}c_{k'}$ with introducing the four Majorana representation, $\sigma_i^{\alpha}=ic_ib_i^{\alpha}$, as in the original work by Kitaev. Suitable fermion parity also required (see SM for fermion parity in the hyperhoneycomb lattice).

	Note that the presence of a thermal transition was previously reported by finding a peak of specific heat ($C_v$) at $T_c /(3J) \sim 6\times 10^{-3}$ for the isotropic exchange interactions \cite{nasu2014vaporization}, which is reproduced in Fig. 1 (b) (blue).  While the specific heat provides useful information on the presence of phase transition, the magnetic character around the critical point still remains unexplored.

	\textit{Dynamic local spin correlation function.\textbf{---}} 
	Let us first consider the local spin correlation function at a site $j$,   
\begin{eqnarray}
	\langle S^z_j(t)S^z_j(0)\rangle =\sum_{m,n}  \frac{e^{-E_m / T}}{\mathcal{Z}}  e^{i(E_m-E_n)t} |\langle m | S_j^z | n \rangle|^2, 
\end{eqnarray}
by using the Lehmann representation with many-body eigenenergy and eigenstate, $E_n$ and $| n \rangle$ at a temperature $T$. The partition function $\mathcal{Z} = \sum_{m} e^{-E_m/T}$ is used. 
To identify the presence and absence of a magnetic order, we  also define its time-average as 
\begin{eqnarray}
D_z \equiv \lim_{T_0 \rightarrow \infty} \frac{1}{T_0} \int_0^{T_0} dt  \, \langle S^z_j(t)S^z_j(0)\rangle. 
\end{eqnarray}
Then, one can easily prove the following two statements (see SM for detailed proof). 
.
\begin{itemize}
\item $D_z\neq 0$ for any spin rotational symmetry broken states such as ferromagnetic or antiferromagnetic states, choosing the $z$ axis as the local magnetic moment direction.
\item $D_z = 0$ indicates the presence of the  spin rotational symmetry by contrapositive.
\end{itemize}
The Fourier transformation of the correlation function is also introduced, 
\begin{eqnarray}
\Psi_j^z(\omega) =  \lim_{\eta \rightarrow 0^+} \int_0^{\infty} \langle S^z_j(t)S^z_j(0)\rangle e^{i (\omega + i \eta )t }.
\end{eqnarray}
Note that a non-zero small broadness parameter $\eta$ is used in Fig. 2.

	\begin{figure}[t]
		\includegraphics[width=4cm, height=4cm]{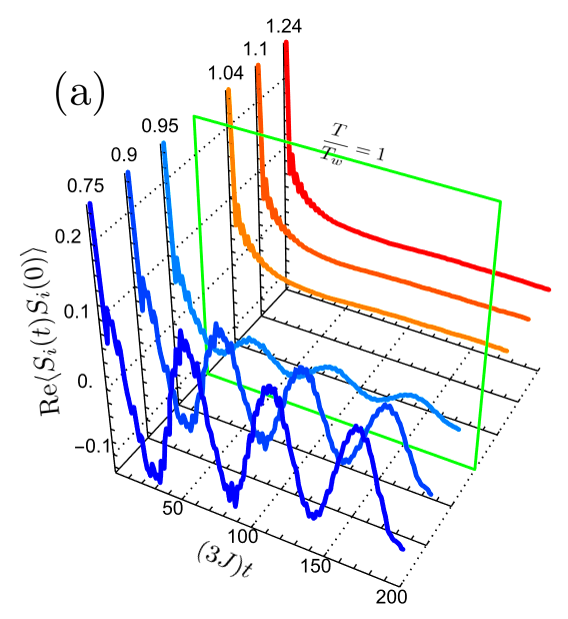}
		\includegraphics[width=4cm, height=4cm]{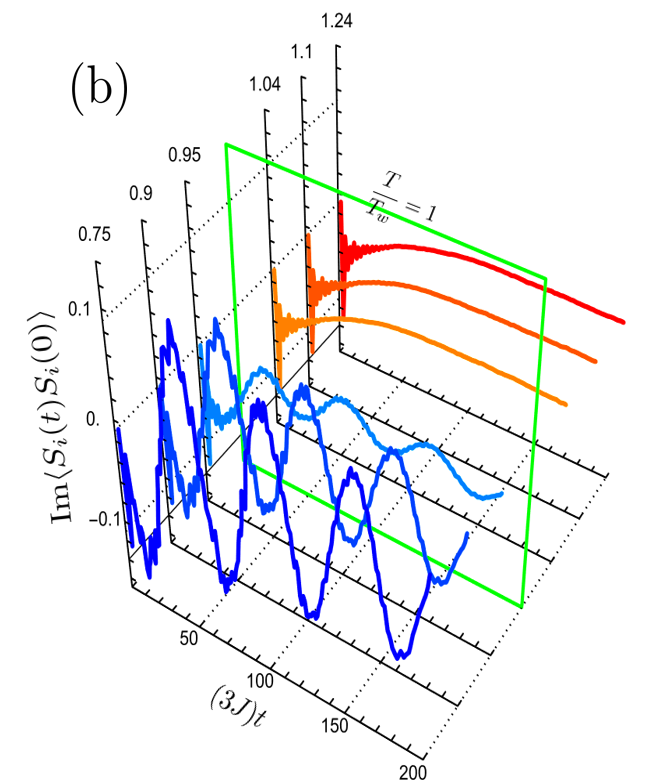}
		\includegraphics[width=4cm, height=4cm]{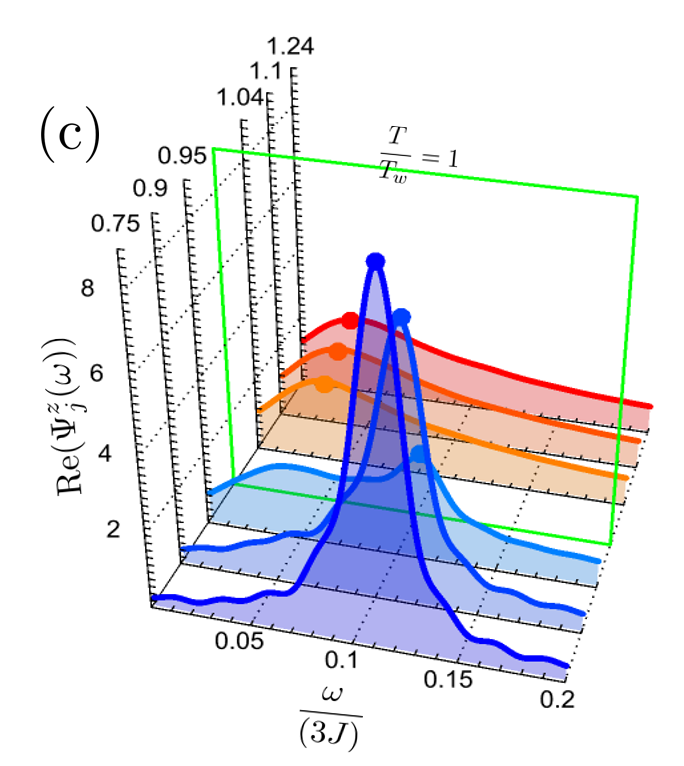}
		\includegraphics[width=4cm, height=4cm]{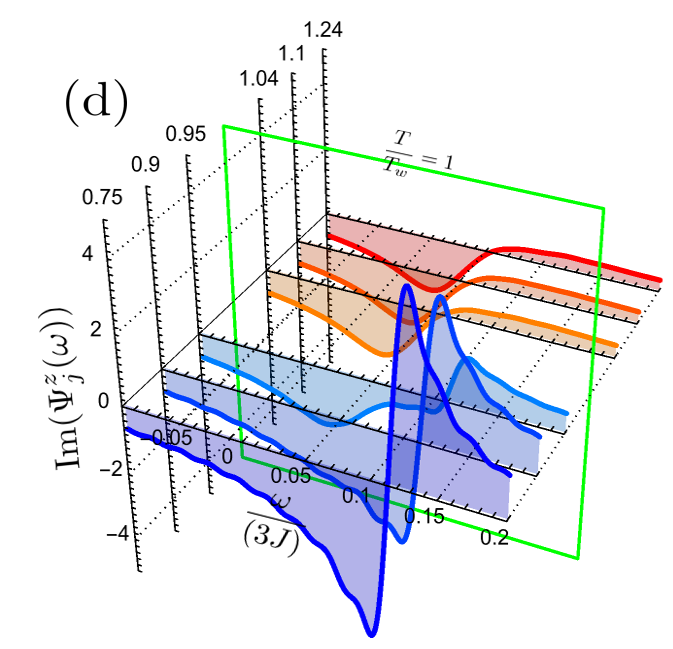}
		
		\caption{Dynamic local spin correlation functions with the periodic boundary condition for $N=4$. Real (a) and Imaginary (b) parts of the correlation functions for six different temperatures.   (c) and (d) are for the real and imaginary parts of the Fourier transformations of the correlation functions.  We use the broadness $\eta=0.09$ for the presentation. The green rectangles are for the transition temperature, $T=T_w$. }
		
		\label{Fig2}
	\end{figure}

With respect to computational cost, it is usually difficult to access the dynamical spin correlation, especially in the time domain. However, the Majorana expression of the Kitaev model makes the calculation possible with practical computational resources.
The dynamical local spin correlation function can be rewritten in terms of the Majorana fermions, and extending the previous results of the pure Kitaev model in two spatial dimensions \cite{udagawa2018vison}, we find the exact expression, 
\begin{eqnarray}
	&& \langle S^z_j(t)S^z_j(0)\rangle=\frac{\sum_{\{W_p\}}\sqrt{\text{det}({1+e^{-(1/T-it)iA}e^{-itiA^{(j)}}})}}{\displaystyle\sum_{\{W_p\}}\sqrt{\text{det}(1+e^{- iA /T})}}  \nonumber  
\end{eqnarray}
where, $A$ is the matrix of Eqn. (2).  $A^{(j)}$ is obtained by flipping two $Z2$ variables at site $j$ from $A$ (see SM for more detailed explanation). One of the key advantages of the exact expression is that Quantum Monte Carlo simulations (QMC) may be utilized to evaluate  the dynamical spin correlation function, without resorting to the use of numerical analytical continuation.
 
Using this, we obtained the correlation function for $N=3,4,5$ when $J=J_x=J_y=J_z=1/3$ for both boundary conditions. While the figure illustrates results with the periodic boundary condition, we have also verified that similar results under  the semi-open boundary condition. We  performed about 6000 Monte Carlo steps for measurements after 3000 steps for thermalization. 
Parallel tempering with 64 replicas, which enables rapid equilibration at lower temperatures, was employed.
We stress that the sign ambiguity of the square roots in $\langle S^z_j(t)S^z_j(0)\rangle$ usually causes serious technical problems, but we succeed to remove the ambiguity by exploiting the Pfaffian properties of $A$ for the first time in this work (see SM for detailed discussion).

\textit{	Dynamic order.\textbf{---}} Our main results are illustrated in Fig.~\ref{Fig2}. We display the dynamical local spin correlation around the critical temperature, $T_w$, with the periodic boundary condition for $N=4$.
In Fig.~\ref{Fig2} (a) and (b), it is clearly shown that the dynamical character of the system sharply changes around $T_w$. Above $T_w$, the spin correlation monotonically decays with characteristic time scale, $3Jt \simeq O(10)$.  
Whereas just below $T_w$, clear oscillatory signature appears.
	The sharp change also appears in frequency domain, where the peak position of $\Psi_j^z(\omega)$ is suddenly shifted from the high temperature value ($\sim0.03$) to the low temperature value ($\sim0.1$), across the critical temperature, planes with the green boundary.

We remark two points of our analysis. 
First, we identify the transition temperature ($T = T_w$) by examining an expectation value of the Wilson loop operator, $\hat{W}$ as defined in \cite{nasu2014vaporization}.
 This choice of Wilson loop operator describes a loop excitation of gauge fields, which can be used to detect a thermal phase transition in a $Z_2$ gauge theory. 
Note that the onset temperature $T_w$ of $ \langle \hat{W} \rangle$ can be different from the peak of specific heat for a finite-size system.
We believe that the two temperatures become identical in thermodynamic limit. To be specific, we determine the transition temperature with the criterion $|\langle \hat{W} \rangle| = 0.03 $, giving   $T_w=6.7\times10^{-3}$ for $N=4$. 
Second, the periodicity of the oscillation in  $\langle S^z_j(t)S^z_j(0)\rangle$ and the peak position of $\Psi_j^z(\omega)$ at low temperatures are insensitive to temperatures in our numerical calculations though their values vary with the system size. 
It means that the oscillation in the low temperature phase may be attributed to a well-defined low energy excitation. 

 To extract further physical meanings of our QMC results,  we employ an approximation focused on the main peak of the spin correlation function to extract further physical insights.
\begin{eqnarray}
	&&{\rm Re}( \langle S^z_j(t)S^z_j(0)\rangle ) \nonumber \\
	&\simeq& A_{\rm stat}(T) + \, A_{\rm dyn}(T) \,e^{-\frac{t}{t_{\phi}(T)} } \cos(\phi(T) t).
\end{eqnarray}
Four parameters ($A_{\rm stat}(T), A_{\rm dyn}(T), t_{\phi}(T), \phi(T)$) are introduced to fit the QMC results. 
The parameter in the cosine function ($\phi(T)$) may be understood as an average of the differences between significant excited states' energy, which become low energy excitations at low temperatures.  
Note that our approximation can be systematically improved by adding additional trigonometric functions in principle. As a sanity check, we find that the QMC results give $A_{\rm stat}(T)=0$ for all cases, equivalent to $ D_z=0$, as it should be.

We extract the temperature dependence of $A_{\rm dyn} (T)$ of our QMC results, which is illustrated in Fig. 1 together with  specific heat calculation and the expectation value of a Wilson loop operator, $\langle \hat{W} \rangle$.  
Surprisingly, the amplitude of the dominant mode, $A_{\rm dyn}(T)$, precisely follows the tendency of $\langle \hat{W} \rangle$ which becomes negligible for $T>T_w$.  
This correspondence implies an unexpected connection between the two quantities concerning the quantum spin liquid phase; a topological quantity $\langle \hat{W} \rangle$, which characterizes the phase theoretically, and the coherency of the dynamics, $A_{\rm dyn}(T)$, representing the experimentally observable quantity of the quantum spin liquid phase. From this correspondence, we argue that $A_{\rm dyn}(T)$ plays a role of an order parameter of the thermal transition at $T_w$ and we call it the dynamic order.

	\begin{figure}[t]
		\includegraphics[width=4cm, height=4cm]{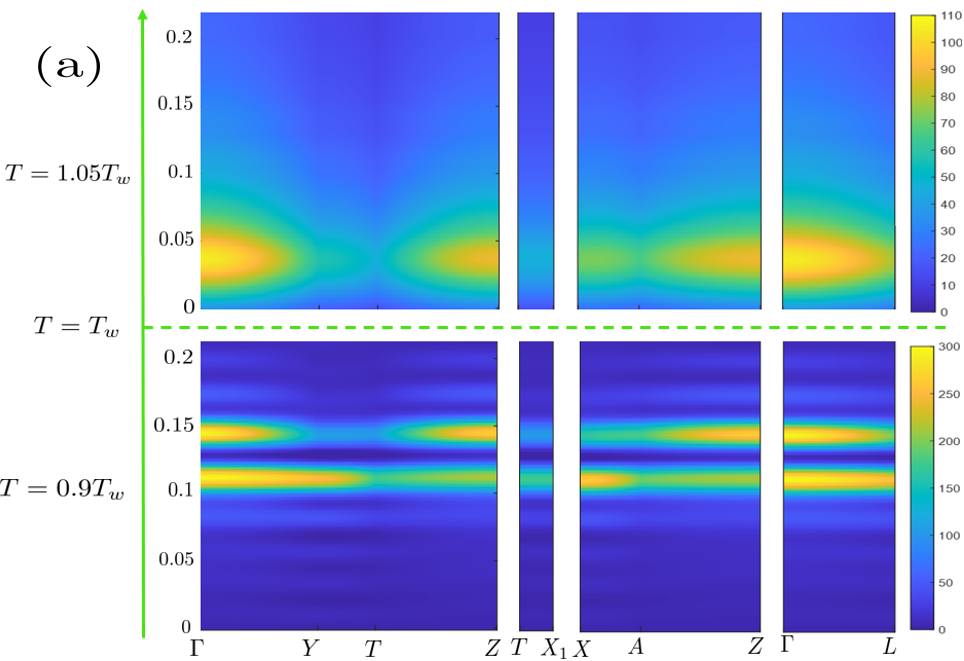}
		\raisebox{0pt}[0pt][0pt]{\vspace{-3cm}\hspace{0.5cm}\includegraphics[width=4cm, height=4cm]{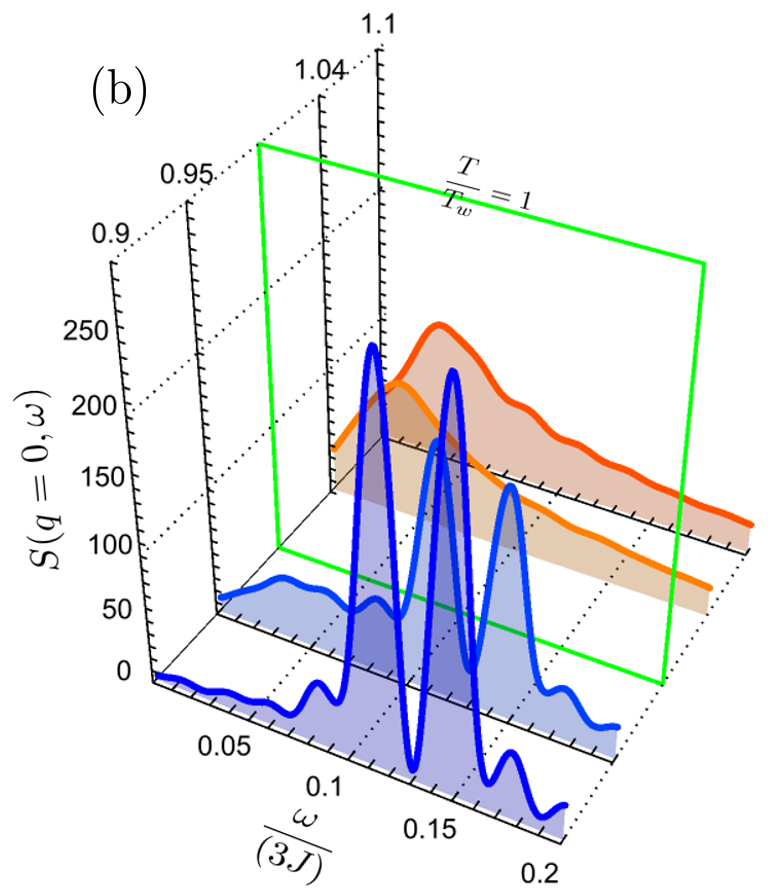}}
		\caption{(a) Numerical results of spin structure factor at 0.9$T_w$ and 1.1$T_w$ with the color scale. The vertical line is for temperature. The sudden changes of the peak positions are clearly shown.  (b) Spin structure factors at $q=0$ are illustrated. The green rectangle is for the transition temperature, $T=T_w$. We use the broadness $\eta=0.01$ for the presentation.}
	\end{figure}

	\textit{Spin structure factor.\textbf{---}} 
	The spin structure factor is defined as
	\begin{eqnarray}
		S(\mathbf{q},\omega)=\frac{1}{N}\displaystyle\sum_{i,j}e^{-i\textbf{q}  (\textbf{r}_i-\textbf{r}_j)}\int_{-\infty}^\infty \mathrm{e}^{i\omega t}\langle \vec{S}_i(t) \cdot \vec{S}_j(0)\rangle\,\mathrm{d}t. \nonumber
	\end{eqnarray}
	Imposing the periodic boundary condition for $N=4$, the spin structure factor is calculated by using our QMC results. The representative results at $T/T_w=0.9$ and $T/T_w=1.05$ along the defined momentum path are illustrated in Fig. 3(a). In Fig. 3(b), its momentum cuts along $\mathbf{q}=0$, $S(0,\omega)$, at four temperatures ($T/T_w=0.9, 0.95, 1.04, 1.1$) are shown.

	The significant differences  between lower and higher temperatures than $T_w$ manifest. For $T=1.1T_w$, the spin structure factor has a broad continuum feature in all the energy scale except the peak around $\omega \sim 0.03$ and $q \sim 0$. Note that the peak position shows the strong temperature dependence. On the other hand, the two more flat dispersion relations around $\omega \sim (0.1, 0.15)$ appears in $T=0.95T_w$ whose positions are temperature independent. Note that our QMC results qualitatively agree with the previous one for the thermodynamic limit at $T=0$ \cite{smith2016majorana}. 
	We suspect that the sharp peaks below $T_w$ is proportional to the dynamic order, $A_{\rm dyn}(T)$.
	
%
%
%
	To check the size dependence, we compare QMC calculations of the local dynamic correlation function below $T_w$ with the parton analysis at $T=0$ following the previous works \cite{smith2016majorana,knolle2014dynamics}. 
	The parton analysis allows us to access a large size of lattice points, in principle the thermodynamic limit ($N \rightarrow \infty$). 
	The results indicate that our QMC results are adiabatically connected to the parton analysis at $T=0$ with larger lattice size points (see Supplementary materials).  
	In Fig. 4, we present the parton analysis with $N=200$ at $T=0$ where the clear oscillations of the real and imaginary parts of local spin correlation functions manifest. We note that the width around the peak frequency is much wider than our QMC results, indicating strong decays even at zero temperature.
	Thus, based on our finite size QMC calculations and the parton analysis at zero temperature, we argue that  the appearance of the peak at $\omega^*$, or equivalently the periodicity of the oscillation, may dynamically characterize the quantum spin liquid phase. 
	
	\begin{figure}[htb!]
		
		\includegraphics[width=4cm, height=4cm]{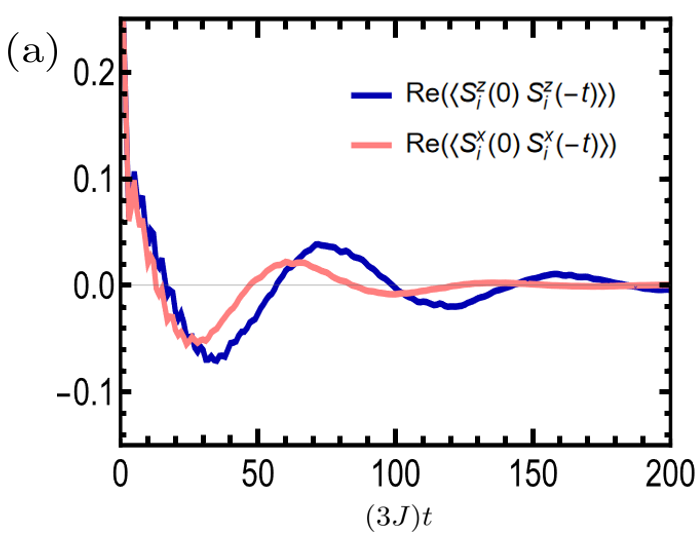}
		\raisebox{1pt}[0pt][0pt]{\hspace{0.5cm}\includegraphics[width=3.93cm, height=3.93cm]{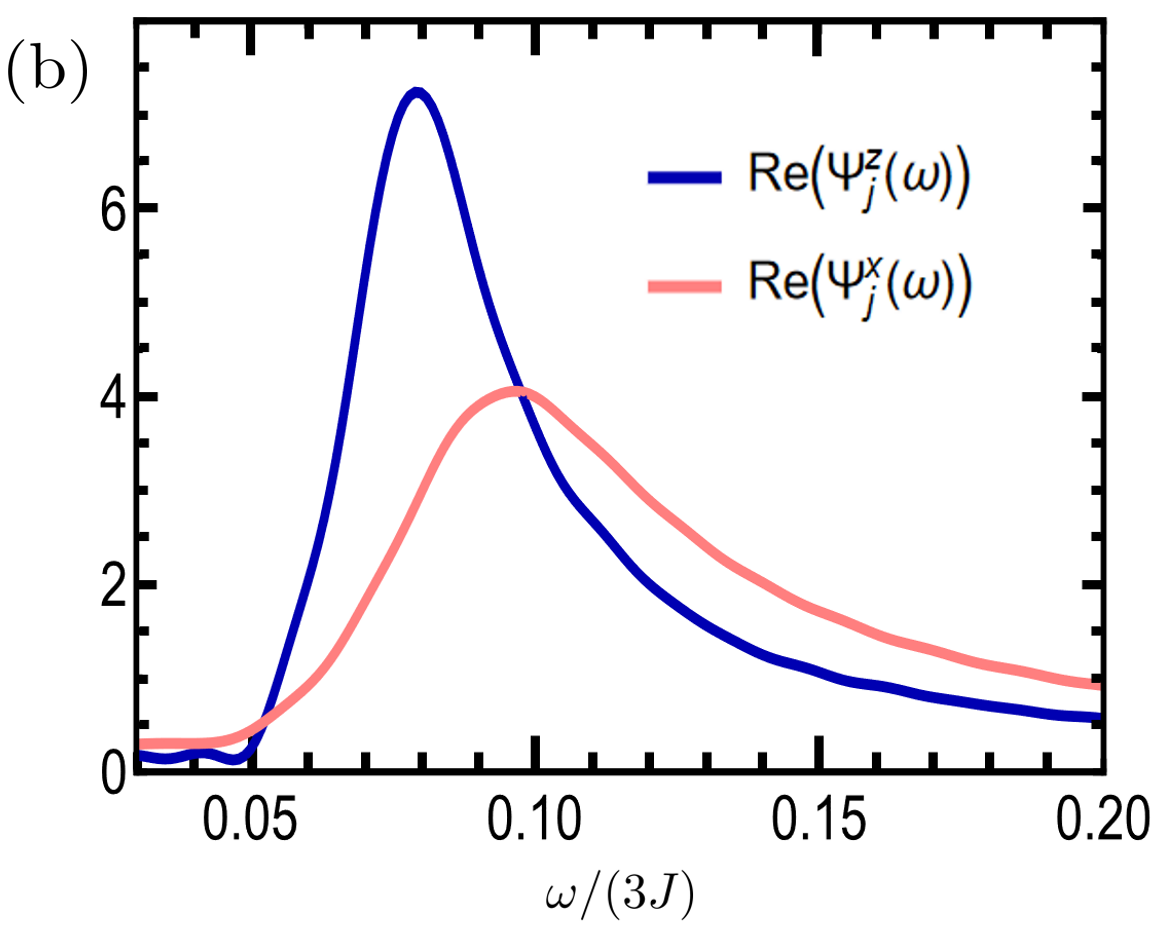}}

		\caption{ (a) Real (blue) and Imaginary (red) parts of the local spin correlation function with $N=200$ at $T=0$. (b) Real (blue) and Imaginary (red) parts of the Fourier transformation of the local spin correlation function with $N=200$ at $T=0$ with $\eta \sim 10^{-3}$.}
	\end{figure}

		\textit{Discussion and Conclusion.\textbf{---}} 
		The physical meaning of the emergent oscillation at $\omega^*$ associated with $A_{\rm dyn}(T) $ needs more discussion. As demonstrated in Fig. 1, the behavior of $A_{\rm dyn}(T)  $ is clearly related to the non-local Wilson loop though $A_{\rm dyn}(T)  $ is from the  dynamic local spin correlation while $\langle \hat{W }\rangle$ is non-local. It is an intriguing open question how the local object $A_{\rm dyn}(T) $ captures the non-local Wilson loop. 
	Our results open a new possibility that the time dynamics of a local spin observable could capture properties of non-local objects, identifying a quantum spin liquid. 
		
	Our theoretical results may be directly applicable to experiments including neutron scattering, dynamic magnetic susceptibility and nuclear magnetic resonance of three dimensional spin liquid candidate materials such as $\beta$-Li$_2$⁢IrO$_3$ \cite{takayama2015hyperhoneycomb,majumder2019anisotropic}.
	We note that such dynamic spin responses are in a spin sector while specific heat measurements capture all thermal degrees of freedom. Thus, our dynamic spin responses are free of trivial phonon contributions. 
	Future research in dynamic spin response functions in three dimensional candidate materials of quantum spin liquids is highly desirable. 
	
	In conclusion, we investigate dynamic signatures of Kitaev's model on the hyper-honeycomb lattice by using quantum Monte Carlo and parton calculations of spin correlation functions. At low temperatures, we uncover the characteristics such as an emergent oscillation of local spin correlation functions of a spin liquid state which disappears at high temperatures, named dynamic order. We propose that a dynamic-order may be applicable to investigate highly entangled quantum many-body systems.

	\emph{Acknowledgement---}  
	M.P. and E.-G.M. were supported by 2021R1A2C4001847, 2022M3H4A1A04074153, National Measurement Standard Services and Technical Services for SME funded by Korea Research Institute of Standards and Science (KRISS – 2024 – GP2024-0015) and the Nano \& Material Technology Development Program through the National Research Foundation of Korea(NRF) funded by Ministry of Science and ICT(RS-2023-00281839). M. U. was supported by JSPS KAKENHI Grant Numbers JP20H05655, JP22H01147, and JP23K22418.

	\bibliographystyle{apsrev}
	\bibliography{./references}


	\clearpage
	
	\addtolength{\oddsidemargin}{-0.75in}
	\addtolength{\evensidemargin}{-0.75in}
	\addtolength{\topmargin}{-0.725in}
	
	\newcommand{\addpage}[1] {
		\begin{figure*}
			\includegraphics[width=8.5in,page=#1]{Supplementary-material_Final.pdf}
		\end{figure*}
	}
	\addpage{1}
	\addpage{2}
	\addpage{3}
	\addpage{4}
	\addpage{5}
	\addpage{6}

\end{document}